**Can we predict individual concentrations of tacrolimus after liver transplantation? Application and tweaking of a published population pharmacokinetic model in clinical practice**

Marie-Astrid Decrocq-Rudler, PharmD, (1, 2)*, Anna H-X. P. Chan-Kwong, PharmD, (1, 2, 3)*, Lucy Meunier (4), MD, Julien Fraisse (5), MSc, José Ursic-Bedoya (4), MD, Sonia Khier (1, 2), PharmD/PhD.

*M-A.D-R and A. H-X. P. C-K. contributed equally to this work.

1. Pharmacokinetic and Modeling Department, School of Pharmacy, Montpellier University, Montpellier, France
2. Probabilities and Statistics Department, Institut Montpellierain Alexander Grothendieck (IMAG), CNRS UMR 5149, Montpellier University, Montpellier, France
3. SMARTc group, Inserm, CNRS, Institut Paoli-Calmettes, CRCM, Aix-Marseille University, Marseille, France
4. Department of Hepato-Gastroenterology and Liver Transplantation, Montpellier University Hospital (Saint Eloi), Montpellier, France
5. Altran-Capgemini, Montpellier, France

Correspondence

Dr. Sonia Khier, Pharmacokinetic Modeling Department, School of Pharmacy, Montpellier University, 15 Avenue Charles Flahault, Montpellier 34000, France (e-mail: sonia.khier@umontpellier.fr).

ORCID ID https://orcid.org/0000-0001-6712-8461

**Conflict of interest declaration:**

JUB: reports travel grants from Astellas outside the submitted work.

Other authors declare no conflict of interest.

**INTRODUCTION**

Tacrolimus (TCL) is the cornerstone of immunosuppression therapy in liver transplantation (LT) for the prevention of acute rejection (1–3). As a member of calcineurin inhibitor's pharmacological class its mechanism of action leads to the inhibition of T-lymphocytes by blocking the transcription of interleukin 2 (4). Absorption rate is variable with an estimated mean bioavailability of 25% (individual bioavailability: 6-43 %). Distribution is limited by a strong erythrocyte and protein (albumin (ALB) and α-1-glycoprotein) binding in plasma. TCL is extensively metabolized by the cytochromes CYP3A4 and CYP3A5, in the gut and liver. Metabolization is therefore the major way of elimination and drug amount excreted in urine is negligible (5,6).

Assessing the optimal individual dose of TCL is difficult in clinical routine because i) the therapeutic range of efficient concentrations is narrow (7,8) ii) there is a strong and significant between-subject pharmacokinetic variability (BSV) leading to a "critical dose" concept - for a drug with a narrow therapeutic range, the concept of a "critical" dose means that each patient requires a different dose of the same drug due to BSV (9,10) and iii) there is a specific between-occasion variability during the first weeks, linked to a non-linear pharmacokinetic on hepatic clearance. TCL apparent clearance (CL/F) increases gradually by 1.8 % per day during the first month post-transplant, which corresponds to the mean time for recovering an entire liver function (11).

Many sources of PK variability have been identified and well described by Venkataramanan et al. (6) on the CL/F and distribution volume. Among them, one of the most meaningful and described in literature, is a CYP3A5 polymorphism (CYP3A5*1) conferring a faster CL/F (12–14). Another one is the post-transplant day or post-operative day (POD) linked to the non-linear CL/F during the first weeks after transplantation (15,16).

Post-transplant therapeutic drug monitoring (TDM) of TCL consists in checking trough concentrations $C_{min}$ to ensure that $C_{min}$ value is close to the therapeutic target concentrations. If not, the dosage must be adjusted according to the $C_{min}$ value and the clinical observations. High level of $C_{min}$ is correlated with side effects or toxicity (17–19), the main one being nephrotoxicity (20). On the contrary, drug underexposure (low area under the curve, AUC) is predictive of acute rejection (21–24).

In addition to usual TDM, a population pharmacokinetic model (PopPK) may be helpful to evaluate an individual dosage: it can be used to define the first dose according to the patient's covariates (*A priori* predictions) and to adapt further doses by assessing individual's pharmacokinetics parameters (Bayesian forecasting) with the monitoring of TCL concentrations. TDM guided by a PopPK model demonstrated to be a better approach to traditional TDM and proved to be useful for decision-making in a clinical setting (25). Many PopPK models of TCL in liver transplanted patients (adults) have been established (26) and one external evaluation of published tacrolimus PopPK models in adult LT patients was conducted (27).

The purpose of the present study was to identify a population PK model for TCL from a literature review and try to improve its predictive performances to estimate TDM concentrations. Refinement of the model was obtained by tweaking the model with "$PRIOR" subroutine of NONMEM software.

**METHODS**

**External cohort and study design**

This retrospective study was approved by the institutional review board (n°2019_IRB_MTP_12-06). Data were collected from adults LT recipients who received TCL therapy at Montpellier University Hospital from January to June 2018 and January to December 2019.

The exclusion criteria were a prolonged stay in intensive care (> 30 days), retransplantation, simultaneous liver-kidney transplantation and patients with missing data. Patient demographic characteristics (including age and gender) and clinical information (aspartate aminotransferase or ASAT, ALB, POD, TCL posology) were collected in the hospital electronic medical record system. As biological assessments were not carried out every day, when a covariate was missing, its value was arbitrarily replaced by the previous day's value.

TCL therapy (Prograf®, immediate release capsules) was initiated at 0.05 to 0.075 mg/kg twice a day (7:15 a.m. in the morning and 6:30 p.m. in the evening) within 48 hours after surgical transplantation. Once the patient reached clinical stability, Prograf® was switched by Advagraf® (TCL extended release form).

Based on the clinical status, acute rejection risk and $C_{min}$ value, an empirical adjustment of TCL amount was done by the clinician to reach or maintain the target range of concentrations, established by the transplant unit: [8-12 ng/mL] during the first four weeks and [5-10 ng/mL] thereafter.

All patients received oral TCL therapy as part of a triple immunosuppressive regimen, which also included mycophenolate mofetil and corticosteroids. For some of them, Basiliximab was administered first, delaying the introduction of TCL. Corticosteroids were administered at 5 mg/kg (Intravenous bolus) in the operating room and then a daily dose of 20 mg of prednisolone *per os*. From Day 7 (POD), a tapering is initiated until stopping at 6 months. Mycophenolate mofetil was started at Day 1 (POD) at a dose of 1g twice daily.

**Sample analysis**

TDM blood samples were collected at 6:00 a.m., three times a week or more if required. Samples were pre-treated to lyse the red blood cells and separate TCL from the proteins. The analysis of the prepared samples was performed by automated antibody conjugated magnetic immunoassay. The lower limit of quantification was 0.86 ng/mL. All the analytical performances of the method are described in Bargnoux et al. (28).

**Literature review**

A literature review was conducted in the PubMed/Medline database, with the following search terms ([tacrolimus] AND [population pharmacokinetic] AND [adult liver transplantation]) from inception to May 31, 2019. Additional relevant studies were manually screened from the identified publications.

Multiple-transplanted cases were not considered in the review and as TCL elimination is hepatic, we considered that the elimination functions of a partial liver transplanted were too different to those of a whole liver transplanted. Therefore, partial liver graft cases were excluded. We also excluded studies involving only prolonged release tablets of TCL and pharmacokinetic drug-drug interaction studies.

We examined only popPK studies (not PK/PD, PBPK) and excluded models developed with another software than NONMEM or if publication did not provide enough information to code the model (structural model, covariates...).

The objective of this review was to select a model which would fit to our clinical context i.e. developed with a population similar to the local population of the Montpellier University Hospital and in accordance with the clinical routine practices.

**Population Pharmacokinetic analysis**

*Dataset*

All the data required for this work were collected retrospectively and merged in an initial dataset set formatted to comply with NONMEM. The initial dataset was split, according to the day of transplantation, into an estimation dataset (70% of the initial dataset) and a prediction dataset (30% of the initial dataset). Management of data and summary of modelling method are detailed in Figure 1.

*Modeling*

Pharmacokinetic analysis was performed using the nonlinear mixed effects modelling software NONMEM® (version 7.4, ICON Development Solutions, Ellicott City, MD, USA). The NONMEM output and post-processing graphs were analyzed and produced using the R software (version 3.6.3, R Foundation for Statistical Computing, http://www.R-project.org).

The selected PopPK model was coded based on the formulas and parameters extracted from original article. The first approach (Figure 1, B) consists in applying the native literature model on prediction dataset and then running a Maximum *A Posteriori* (MAP) Bayesian analysis. The second approach (Figure 1, A) consisted in adapting the native literature model to the target population with the $PRIOR approach. The native literature model was run on the estimation dataset with the $PRIOR subroutine (NONMEM®). First, we simply used full informative priors: the weight of the prior and the possibility for the model to deviate from the informed prior value depended on the precision of the reference parameters, as quantified by their standard error. The lower the standard error, the heavier the parameter in the prior information, the more restricted the possibility to deviate from its value. The new estimated parameters composed the tweaked model F. Then, we optimized the weight of the priors to minimize the influence of prior information with a correct estimation on the estimation dataset, as recommended by Chan Kwong *et al.* (29). The model obtained was called the tweaked model O. The tweaked F and O models were used to estimate predictions by MAP Bayesian analysis (MAXEVAL=0) on the prediction dataset.

Both *a priori* and Bayesian predictions were computed. *A priori* predictions were computed using covariates only (all observations set at MDV=1). Bayesian predictions of the second and next concentrations were based on the previous observation(s) of each patient (subsequent observations were set at MDV=1).

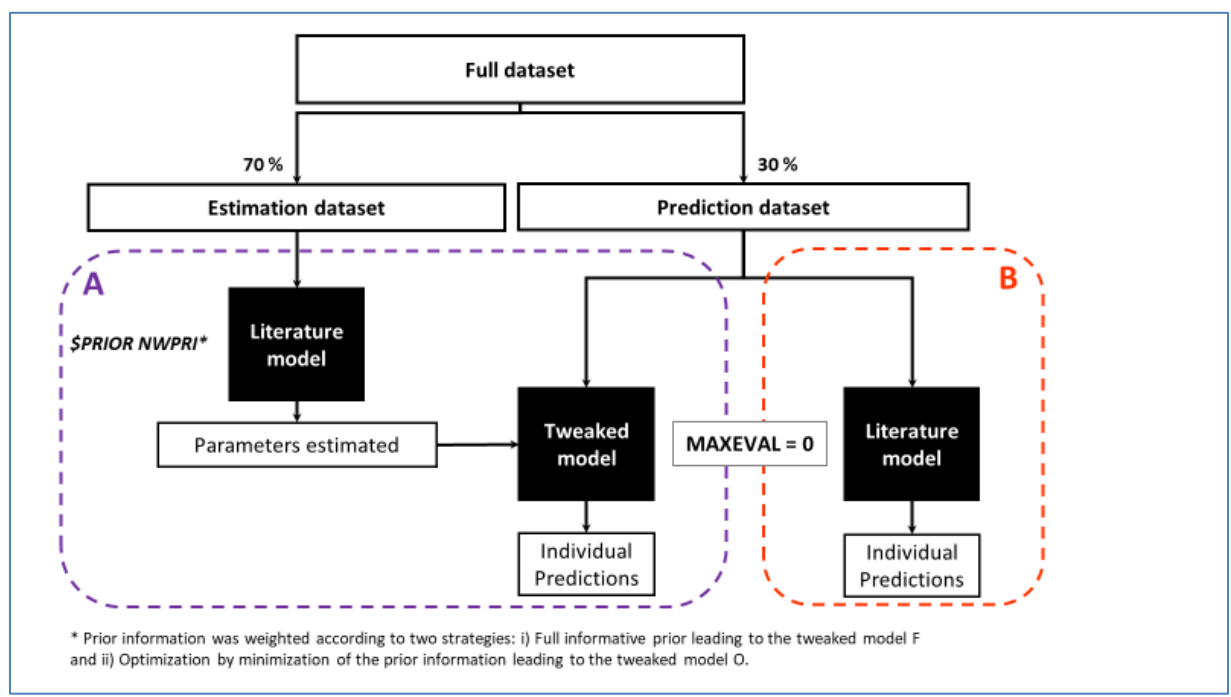

* Prior information was weighted according to two strategies: i) Full informative prior leading to the tweaked model F and ii) Optimization by minimization of the prior information leading to the tweaked model O.

Figure 1. Modeling methodology. A: $PRIOR subroutine approach, B: classical approach with literature model.

*External evaluation of predictability*

Prediction error (PE) (30) were calculated and used to assess the final model's predictive performance in terms of precision and bias (Eq. (1)).

$$PE_{ij} \ (\%) = \frac{(I)PRED - OBS}{OBS} * 100 \hspace{2cm} (1)$$

PRED: Population predicted concentration

IPRED: Individual predicted concentration

OBS: Observed concentration

The median PE% (MDPE) and median absolute PE% (MDAPE) were applied to evaluate the bias and imprecision of the predictive performance, respectively (31).

$$MDPE \ (\%) = median(PE_{ij}, j = 1, \dots N_i)$$

$$MDAPE \ (\%) = median(|PE_{ij}|, j = 1, \dots N_i)$$

The MDAPE of literature and tweaked models were statistically compared with a pairwise one-sided paired Wilcoxon sign-ranked test (Bonferroni correction).

As a combination predictor of both accuracy and precision, $F_{20}$ (proportion of PE% within ±20%) and $F_{30}$ (proportion of PE% within ±30%) were also calculated. The predictive performance of a candidate model was considered satisfactory and clinically acceptable if the following criteria were met: - 20% ≤ MDPE ≤ 20%, MDAPE ≤ 30%, $F_{20}$ ≥ 35% and $F_{30}$ ≥ 50%. These criteria of acceptability were previously used for external evaluation of published population pharmacokinetic models for tacrolimus in adult liver transplant recipients (27). A PE% within ±20% allows to predict a concentration within the target range: 10 ±2 ng/mL during the first four weeks and 7.5 ±1.5 ng/mL after the first month.

**RESULTS**

**Demographic characteristics and clinical data of external cohort**

109 LT patients were enrolled in this study. 40 were excluded due to retransplantation, long stay in intensive care unit or incomplete information. The final dataset included 851 observations from 79 patients. For PopPK analysis, the final dataset was divided into an estimation dataset (70% of the initial dataset, 561 observations from 55 patients), and a prediction dataset (30% of the initial dataset, 290 observations from 24 patients). Demographic and clinical characteristics are summarized in Table 1.

Hepatocellular carcinoma (31.6%) and alcohol-associated cirrhosis (31.6%) were the main LT indications. The final dataset included 851 concentrations ($C_{min}$, obtained from TDM) with an average of 11 observed concentrations per patient. The scatterplot of observed concentrations [Min = 1.2 ng/ml,  Max = 27.6 ng/ml]  (SupData_Document1) underlines concentration variability during the first month (n=1 to 4 weeks) following the transplant.

**TABLE 1.** Demographic, Clinical, Biological, and Pharmacologic Characteristics

| | Full Data set | Estimation Data set | Prediction Data set |
|---|---|---|---|
| No. of patients, n | 79 | 55 | 24 |
| Gender, n (%) | | | |
|   Female | 29 (37%) | 19 (34.5%) | 9 (37.5%) |
|   Male | 50 (63%) | 36 (65.5%) | 15 (62.5%) |
| Age (yr), mean (SD) [min–max] | 56.2 (10.5) [19–71] | 55.2 (11.2) [19–69] | 58.6 (8.9) [34–71] |
| Weight (kg), mean (SD) [min–max] | 76.4 (16.1) [51–130] | 75.8 (19.6) [51–130] | 77.8 (12.6) [60–97] |
| Clinical background, n (%) | | | |
|   Alcohol-associated cirrhosis liver tumor: | 25 (31.6%) | 15 (27.3%) | 10 (41.7%) |
|     Hepatocellular carcinoma | 25 (31.6%) | 16 (29%) | 9 (37.5%) |
|     Benign liver tumor | 1 (1.2%) | 1 (1.8%) | |
|   Cirrhosis hepatitis C acute liver failure: | 5 (6.3%) | 3 (5.4%) | 2 (8.3%) |
|     Fulminant hepatitis | 1 (1.2%) | 1 (1.8%) | |
|     Acute liver failure (other causes) | 4 (5.1%) | 4 (7.3%) | |
|   Biliary pathology: | | | |
|     Primary sclerosing cholangitis | 2 (2.5%) | 2 (3.6%) | |
|     Secondary biliary cirrhosis | 2 (2.5%) | 2 (3.6%) | 2 (8.3%) |
|   Others causes: | | | |
|     Cirrhosis other known causes | 2 (2.5%) | 2 (3.6%) | |
|     Cirrhosis of autoimmune origin | 4 (5.1%) | 4 (7.3%) | |
|     NASH cirrhosis | 2 (2.5%) | 1 (1.8%) | 1 (4.2%) |
|     Budd Chiarri | 1 (1.2%) | 1 (1.8%) | |
|     Hepatic polycystosis | 1 (1.2%) | 2 (3.6%) | |
|     Other specified cause | 2 (2.5%) | 2 (3.6%) | |
|     Other undetermined cause | 2 (2.5%) | 2 (3.6%) | |
| Biological levels, mean (SD), [min–max] | | | |
|   MELD (model of End-Stage liver Disease) | 21.9 (10.9) [6–40] | 23.3 (11.3) [6–40] | 19.3 (10.2) [6–36] |
|   ASAT (UI/L) | 56.7 (132) [5–4711] | 56.1 (127) [5–4229] | 58.1 (143) [9–4711] |
|   Albumin (g/L) | 32.4 (4.46) [15–49] | 32.5 (4.34) [20–45] | 32.1 (4.68) [15–49] |
| Tacrolimus treatment | | | |
|   POD (d), mean (SD), [min–max] | 18.3 (12.7) [1–77] | 17.7 (11.9) [1–74] | 19.8 (14.3) [1–77] |
|   Tacrolimus daily dose (mg/d), mean (SD) [min–max] | 4.4 (2.8) [0–19] | 4.7 (2.7) [0–16] | 4.1 (2.8) [0–19] |
|   Tacrolimus daily dose (mg/kg/d), [min–max] | [0–0.28] | [0–0.21] | [0–0.28] |
| No. of observed concentrations, n | 851 | 561 | 290 |
| No. of observed concentrations per patient, mean [min–max] | 10.7 [2–28] | 10.5 [4–28] | 11.3 [4–27] |
| Tacrolimus initiation (POD), mean | 2.8 | 2.9 | 2.6 |
| Length of stay at hospital (d), mean [min–max] | 28.9 [6–77] | 28.2 [6–74] | 30 [13–77] |

NASH, nonalcoholic steatohepatitis.

Figure 2 shows the distribution of observed concentrations according to POD. The local target trough concentrations are [8-12 ng/mL] during the first month and [5-10 ng/mL] the following weeks. During the first month, less than half of patients are within the therapeutic range, with a tendency of underexposure. This trend decreases over time along with an increase of patients in the therapeutic range. After one month, the patients are rather well balanced or sometimes overexposed.

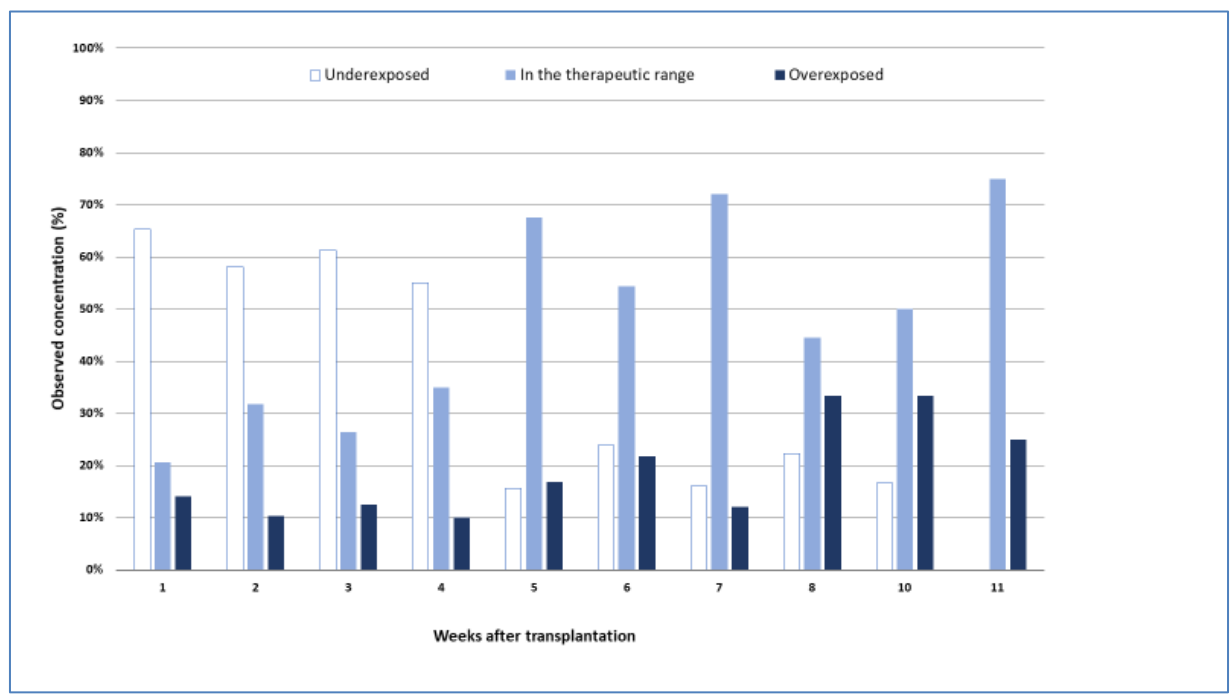

Figure 2. Observed concentration ($C_{min}$) distribution according to week post-transplant.

**Review of published PopPK models**

Details of the literature model selection process are provided in SupData_Document2. The literature review led to consider 70 articles after the initial screening, among them 56 were directly excluded in a first step. 4 articles found in the references of the 70 initial articles were added to our review, leading to 18 models (15,16,32–47) potentially interesting for our study. In a second step, the 18 models were examined and exclusion criteria were applied. Models involving partial liver graft (16,41–43), not in accordance with a NONMEM popPK approach (33–35) or with extra-routine covariates (15,36,37,39,40) were excluded. The final criterion to select the model was the similarity between our patient population and the patient population of the previous models. Indeed, BSV on pharmacokinetics could be linked to population ethnic origins (e.g. P-glycoprotein and CYP allelic variants). That is why the studies conducted in non-European countries were excluded (38,44–46) and finally the model built with a French population was selected (32). The final selected model was developed in 2005 with 37 patients. The aim was to characterize elimination, therefore absorption parameters were not estimated. The covariates included in the model were POD, ALB and ASAT, all implemented on CL/F. The feature of this model is the demonstration of a sigmoid relationship between CL/F and POD. This clearance function is defined by the equation (2) which requires the estimation of additional parameters: $CL_{max}$ (maximal clearance value), $TCL_{50}$ (time needed to obtain 50% of $CL_{max}$) and $\gamma$ (sigmoidicity coefficient).

$$CL = \frac{Cl_{max} \cdot POD^{\gamma}}{TCL_{50}^{\gamma} + POD^{\gamma}} \qquad (2)$$

**Modeling**

The parameters of the literature model were successfully tweaked and optimized on the estimation dataset. The estimation dataset was informative enough to estimate the tweaked model O without covariates and with uninformative priors on $\omega^2$ TCL50. NONMEM scripts and a table summarizing the parameter values and standard error for each model are available in Supplemental Digital Content (SupData Documents 3 and 4).

**External evaluation of predictability**

The results of predicted-based diagnostics are provided in Figure 3 and Table 2. Figure 3A is the boxplot of PE from the literature model and the tweaked models with *a priori* predicted concentrations ($N_{observation}$ = 0) *versus* Bayesian prediction ($N_{observation(s)}$ = 1 to 9). Perfect box plots should fall within the two dashed lines, median around zero. The MDAPE were statistically significantly lower with tweaked than with literature model for *a priori* predictions (p=4.98$^{-15}$ for tweaked F, p=6.06$^{-5}$ for tweaked O).

**TABLE 2.** Criteria of Predictive Performance: Top Table, at Each Step of TDM; Bottom Table, According to the Type of Predictions

| $N_{obs}$ | Type of Model | $N_i$ | Total $N_{PRED}$ | $N_{PRED}$ per patient [Min-Max] | MDPE (%) | MDAPE (%) | F20 (%) | F30 (%) |
|---|---|---|---|---|---|---|---|---|
| 0 | Literature | 24 | 290 | [5–28] | *−41.0* | 45.4 | 16.2 | 28.6 |
| 0 | Literature Cai et al. | 0 | NA | NA | *−44* | *−46* | *18* | *29* |
| 0 | Tweaked F | 24 | 290 | [5–28] | −28.5 | 38.0 | 26.6 | 40.7 |
| 0 | Tweaked O | 24 | 290 | [5–28] | **−8.73** | 39.2 | 34.5 | 41.7 |
| 1 | Literature | 24 | 24 | 1 | 1.01 | *36.1* | *33.3* | *33.3* |
| 1 | Tweaked F | 24 | 24 | 1 | 1.60 | *48.7* | *25.0* | 41.7 |
| 1 | Tweaked O | 24 | 24 | 1 | 10.6 | 38.0 | **37.5** | *41.7* |
| 2 | Literature | 24 | 24 | 1 | −10.4 | 15.2 | 54.2 | 62.5 |
| 2 | Tweaked F | 24 | 24 | 1 | 5.81 | 23.8 | 41.7 | 62.5 |
| 2 | Tweaked O | 24 | 24 | 1 | *21.3* | 24.5 | *25.0* | 58.3 |
| 3 | Literature | 24 | 24 | 1 | −16.2 | 22.3 | 45.8 | 58.3 |
| 3 | Tweaked F | 24 | 24 | 1 | 0.701 | 24.0 | 37.5 | 62.5 |
| 3 | Tweaked O | 24 | 24 | 1 | 8.92 | 23.9 | 45.8 | 62.5 |
| 4 | Literature | 24 | 24 | 1 | *−20.3* | 24.8 | 37.5 | 58.3 |
| 4 | Tweaked F | 24 | 24 | 1 | −10.7 | 21.7 | 50.0 | 70.8 |
| 4 | Tweaked O | 24 | 24 | 1 | −6.86 | 19.3 | 50.0 | 79.2 |
| 5 | Literature | 23 | 45 | [1–2] | −1.70 | 20.0 | 51.1 | 68.9 |
| 5 | Tweaked F | 23 | 45 | [1–2] | 8.40 | 22.8 | 40.0 | 64.4 |
| 5 | Tweaked O | 23 | 45 | [1–2] | 10.6 | 23.2 | 35.6 | 64.4 |
| 7 | Literature | 19 | 35 | [1–2] | −8.70 | 28.1 | 37.1 | 57.1 |
| 7 | Tweaked F | 19 | 35 | [1–2] | −1.60 | 25.4 | 42.9 | 68.6 |
| 7 | Tweaked O | 19 | 35 | [1–2] | 0.333 | 25.3 | 40.0 | 62.9 |
| 9 | Literature | 14 | 90 | [1–19] | −8.30 | *31.1* | *33.3* | 48.9 |
| 9 | Tweaked F | 14 | 90 | [1–19] | −7.40 | 27.6 | 38.9 | 53.3 |
| 9 | Tweaked O | 14 | 90 | [1–19] | −5.69 | 28.3 | 40.0 | 52.2 |

| $N_{obs}$ | Type of model / Type of predictions | $N_i$ | Total $N_{PRED}$ | $N_{PRED}$ per Patient [Min-Max] | MDPE (%) | MDAPE (%) | F20 (%) | F30 (%) |
|---|---|---|---|---|---|---|---|---|
| 0 | Literature a priori | 24 | 290 | [5–28] | *−41.0* | 45.4 | 16.2 | 28.6 |
| 0 | Tweaked F a priori | 24 | 290 | [5–28] | −28.5 | 38.0 | 26.6 | 40.7 |
| 0 | Tweaked O a priori | 24 | 290 | [5–28] | **−8.73** | 39.2 | 34.5 | 41.7 |
| [1–9] | Literature Bayesian | 24 | 266 | [4;27] | −8.35 | 27.3 | 40.2 | 54.9 |
| [1–9] | Tweaked F Bayesian | 24 | 266 | [4;27] | −2.1 | 24.0 | 39.5 | 59.4 |
| [1–9] | Tweaked O Bayesian | 24 | 266 | [4;27] | 2.47 | 24.4 | 39.1 | 58.6 |

$N_{obs}$, number of observations in Bayesian forecasting; $N_i$, number of patients from the "prediction data set"; $N_{PRED}$, number of predictions; NA, not available; MDPE, median prediction error.

The predictive performance of a candidate model was considered satisfactory and clinically acceptable if the following criteria were met: $−20\% \leq MDPE \leq 20\%$, MDAPE $\leq 30\%$, $F_{20} \geq 35\%$, and $F_{30} \geq 50\%$. Italic values are outside the threshold. Bold values are the values that are within the acceptance limits of the quality criterion for "tweaked O model" whereas these values were out of limits for the "tweaked F model"

Predictive performance at each step of TDM are presented in Figure 3B. PE% and other predictive criteria were performed for each concentration and with the previous sample(s) obtained from TDM. Details of criteria of predictive performance showed a global trend to a better performance of tweaked models for bias (MDPE closer to zero) and a similar imprecision (Table 2). A priori predictions with the selected literature model were unsatisfactory. However, the tweaked models showed a better predictive performance criteria : a -41% bias for literature model versus -28.5% and -8.73% for tweaked F and O respectively (threshold between -20 and +20 %) and a 45.4% imprecision for literature model versus 38.0% and 39.2% for tweaked F and O respectively (threshold < 30 %).

The Bayesian forecasting with previous information (observed concentrations) improved bias and imprecision, respectively as from the second prediction (Nobs ≥1 previous observed concentration) and as from the third prediction (Nobs ≥2 previous observed concentrations). Whatever the forecasting state, the tweaked models tend to obtain better results.

**Figure 3.** Box plots prediction error (PE %) for literature model (i: orange box plots) and tweaked model F (ii: purple box plots) and tweaked model O (iii: grey box plots) based on the prediction data set. The two dotted lines represent the threshold of %PE 20 % and 30 %. $N_{obs}$ = Number of previous observed concentration(s)/patient considered for Bayesian forecasting. $N_{PRED}$ = Total number of individual predicted concentration(s) displayed in the boxplot.

# p-values are calculated with pairwise one-sided paired Wilcoxon sign-ranked test (Bonferroni correction).

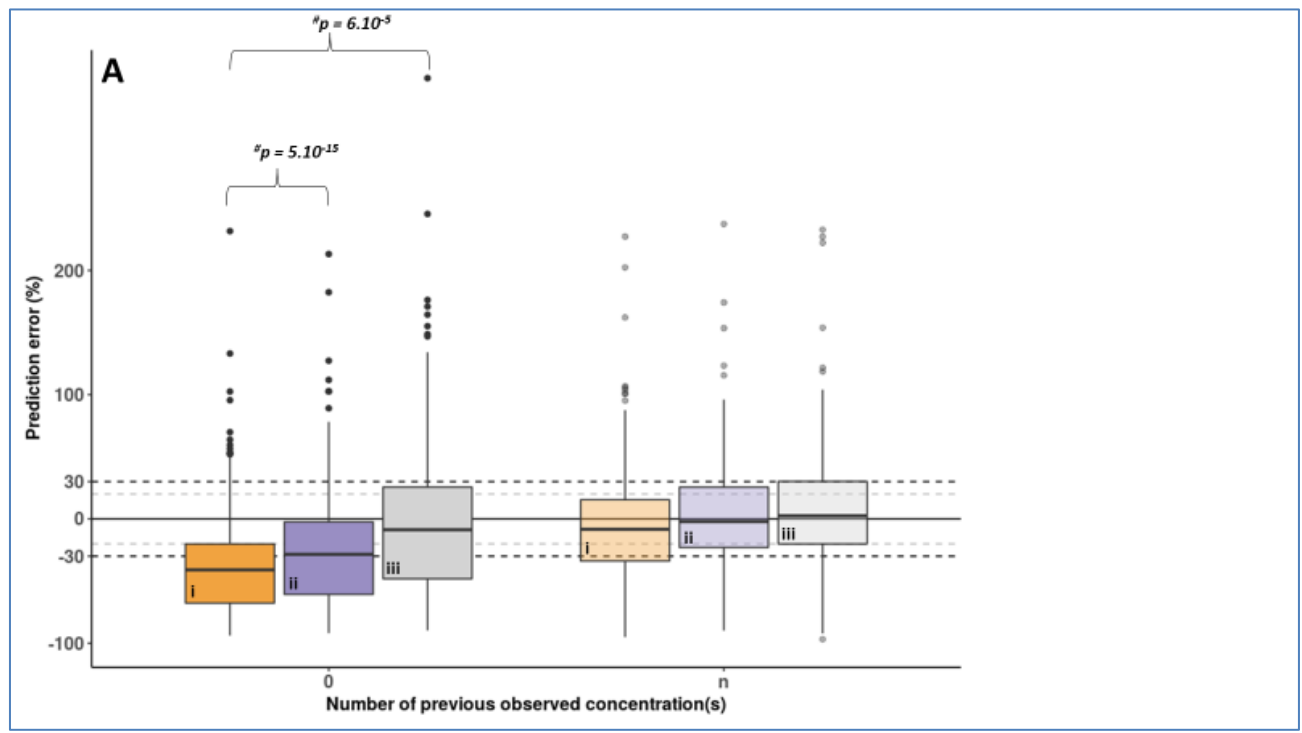

Figure 3A.
Boxplots "0": $N_{obs}$ = 0, $N_{PRED}$ = 290 (5 to 28 per patient, 24 patients).
Boxplots "n": $N_{obs}$ = 1 to 9, $N_{PRED}$ = 266 (1 to 27 per patient, 24 patients).

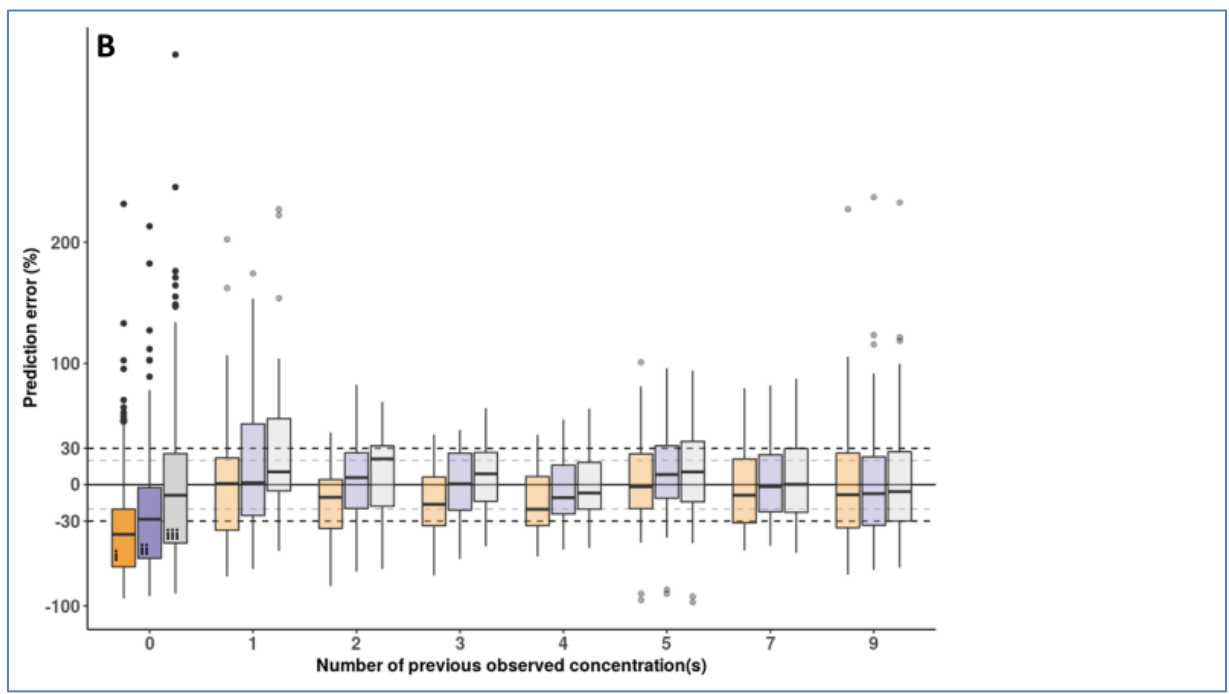

Figure 3B.
Boxplot "1 to 4": $N_{obs}$ = 1 to 4, $N_{PRED}$ = 24 (1 per patient, 24 patients).
Boxplot "5": $N_{obs}$ = 5, $N_{PRED}$ = 45 (1 to 2 per patient, 23 patients).
Boxplot "7": $N_{obs}$ = 7, $N_{PRED}$ = 35 (1 to 2 per patient, 19 patients).
Boxplot "9": $N_{obs}$ = 9, $N_{PRED}$ = 90 (1 to 19 per patient, 14 patients).

## DISCUSSION

TCL is a key drug of transplant rejection and given the clinical issues (high BSV and narrow therapeutic range) this drug requires individualized dosage prescription. To adjust prescriptions, physicians exploit both the available clinical information and TDM ($C_{min}$). It is then up to the physician, based on his experience, to adjust the dosage over the days. The stakes are high since troubles finding the optimal dosage can lead to increased risk of rejection and length of stay. Descriptive statistics on our TDM values confirm the difficulties for physicians to reach the therapeutic range concentration: one month is needed to get at least 50 % of the concentrations in the therapeutic range concentrations.

PopPK model-guided dose adjustment is a tool that helps to manage this issue, by estimating individual patient pharmacokinetic parameters and thereby evaluating a dosage that matches their parameters. To do this, it is possible to use models from literature. We identified 18 PopPK models in literature for TCL with a first total LT, but no one can assert that these models could be used directly on an external cohort of patients, with good predictability. Cai *et al.* (27) published the first comprehensive external evaluation of published tacrolimus PopPK models in adult LT patients using an independent dataset (Chinese population), which was prospectively collected from routine TDM. They concluded that all models had unsatisfactory performances. Nonetheless one model was superior to the other, the Zhang's model (46), based on Asian patients. Globally in Cai *et al.* study, lower MDPE and MDAPE are observed with the Asian models compared to the European models which underlines the importance of ethnicity in the choice of a PopPK literature model.

Our work is in line with the study by Cai *et al.* but, given the unsatisfactory predictive performances of all literature models, we chose another strategy. Firstly we carefully selected the most appropriate PopPK model for dosage prediction, the best suited to our clinical practice and our patients, and secondly we tested its predictive capacities. For example, we excluded Asian population because in Asia HBV and HCV cirrhosis indications are overrepresented, whereas in Europe the main indications are hepatocellular carcinoma and alcohol-associated cirrhosis. Indeed, HCV infection is associated with auto-antibody production directed against CYP3A4, which influences the pharmacokinetics of TCL (48).

Our results demonstrate that the PopPK model selected from literature (Antignac's model (32)) allows satisfactory predictions when at least two previous individual observations ($N_{obs} \geq 2$) inform the model, *i.e.* the third and subsequent concentrations are accurately predicted. The third sample was collected in average one week after the first TCL administration. Thus, the use of this model to guide dose adjustment would really decrease the time to reach the therapeutic range concentration, roughly from one month to one week. In contrast, Cai et al. needed at least four previous concentrations ($N_{obs} \geq 4$) to correctly predict the concentrations of their cohort with the model by Antignac et al. (32). This highlights the importance to target the literature model, eventually with similarities between the population that was used to build the literature model and the external dataset, before testing predictive performance. A model could be poorly predictive for a population and be useful for another situation.

Considering the poor external predictive ability of literature models of TCL reported by Cai *et al.*, we also explored if the $PRIOR function could help to better fit the pharmacokinetics of our population, by tweaking the PopPK parameters with a subset of our data. To the best of our knowledge, this is the first study combining external evaluation of published PopPK of TCL with $PRIOR approach.

In our case where we had collected a quite rich estimation dataset, 561 concentrations from 55 patients, we decided to test on the one hand the simple method of using informative prior on all parameters, resulting in the so-called "tweaked model F", and on the other hand the more time-consuming approach of optimizing the model, resulting in the so-called "tweaked model O". The optimization of the model demonstrated that our estimation data were informative enough to estimate all model parameters except the interindividual variability of one of the characteristic parameters of the apparent clearance ($\omega^2$ TCL50) and the covariates ASAT and ALB. Compared to the literature model, which was built on 728 concentrations from 37 LT patients, our tweaked models had larger interindividual variabilities: this might be linked to the larger number of patients in our estimation dataset. Moreover, the covariate effects of the literature model were not captured with our estimation data. It must be noticed that the covariate effects of the literature model were already not precisely estimated (high relative standard error), as reported in the article by Antignac (32). In terms of predictive performances, we outlined that simply tweaking the model with all informative priors (tweaked F) was comparable to optimizing the model (tweaked O).

Without Bayesian forecasting (predictions *a priori)*, the literature model did not have an acceptable predictive ability. This is in accordance with the results obtained by Cai et al. The tweaked models provided better predictions than the literature model, without, however, reaching the acceptability criteria. Nevertheless, the $PRIOR approach seems to be interesting to obtain better predictions *a priori* during the critical period (first month) in this clinical context. Hence, published PopPK models could be used for individualization of posology with Bayesian forecasting. Without the implementation of Bayesian forecasting, *i.e.* for *a priori* predictions, it is preferable to tweak the model to the target population, using collected data from some previous patients. The PRIOR approach is of particular interest given the struggle to build a model for a specific population in clinical practice, due to the sparseness of the available data.

The empirical choice of the model, although reasoned, could appear as a limitation of our study. Indeed, the different models used by Cai *et al.* had very different predictive performances that were not all explained by population divergences. Moreover, our choice of literature model was conditioned by the non-availability of pharmacogenetic analysis in routine. Pharmacogenetic covariates could improve the predictive capacity.

CONCLUSION

In conclusion, the predictive performance of the selected literature PK model in liver transplanted patients was correct with Bayesian forecasting, but insufficient for *a priori* predictions. To improve the predictive performance, tweaking the literature model with the $PRIOR approach allows to obtain better predictions during the critical period (first month) and after. These results open up opportunities to generalize the use of previous models in clinical practice. External evaluation in the population of interest is a necessary step before defining new dose adaptation. Script (input code) pooling could ease both the external validation and the implementation of PopPK models in TDM. Up to now, it is rare to find a complete script of a model in an original article, or the script communicated by the authors.

**Acknowledgment**

*We thank Guilhem Darche, David Marchionni and David Fabre from Sanofi R&D Montpellier for tips and help for data management and the provision of NONMEM licence. Thanks to Mathieu Morell from the hospital of Montpellier for data extraction.*

**Conflict of interest declaration:**

JUB: reports travel grants from Astellas outside the submitted work.

Other authors declare no conflict of interest.